\documentclass[%
 reprint,
 amsmath,amssymb,
 aps,
]{revtex4-1}
\usepackage{color}
\usepackage{graphicx}
\usepackage{dcolumn}
\usepackage{bm}

\newcommand{\edit}[1]{\textcolor{black}{#1}}
\newcommand{\rev}[1]{\textcolor{black}{#1}}

\begin{document}

\preprint{APS/123-QED}

\title{Deformation and shape of flexible, microscale helices in viscous flow}

\author{Jonathan T. Pham,$^{1,2}$ Alexander Morozov,$^{3}$ Alfred J. Crosby,$^{1}$ Anke Lindner,$^{2}$ and Olivia du Roure$^{2}$}
\affiliation{$^{1}$ Polymer Science and Engineering Department, University of Massachusetts Amherst, 120 Governors Drive, Amherst, MA 01003, USA. \\ $^{2}$PMMH-ESPCI-ParisTech, UMR 7636 CNRS-ESPCI, Universit\'e Pierre et Marie Curie, Universit\'e Paris Diderot, 10 rue Vauquelin, 75005 Paris, France \\ $^{3}$ SUPA, School of Physics and Astronomy, University of Edinburgh, Peter Guthrie Tait Road, Edinburgh EH9 3FD, UK}
\email{crosby@mail.pse.umass.edu \\ anke.lindner@espci.fr \\ olivia.duroure@espci.fr}
\altaffiliation{\\$^{\mp}$Present address: Max-Planck-Institut f\"ur Polymerforschung, Ackermannweg 10, 55128 Mainz, Germany}



\date{\today}

\begin{abstract}
We examine experimentally the deformation of flexible, microscale helical ribbons with nanoscale thickness subject to viscous flow in a microfluidic channel.  Two aspects of flexible microhelices are quantified: the overall shape of the helix and the viscous frictional properties. The frictional coefficients determined by our experiments are consistent with calculated values in the context of resistive force theory. Deformation of helices by viscous flow is well-described by non-linear finite extensibility. Under distributed loading, the pitch distribution is non-uniform and from this, we identify both linear and non-linear behavior along the contour length of a single helix.  Moreover, flexible helices are found to display reversible global to local helical transitions at high flow rate.
\end{abstract}

\maketitle


Helices have captured the fascination of many for centuries, from Darwin's observation of plant tendrils \cite{Darwin1875} to a child's play with a toy Slinky.  Beyond curiosity, the interaction of small helices with fluids is particularly important because of its relevance to both fundamental science \cite{Wada2009,Teran2010,Rodenborn2013,Liu2011,Lauga2009,Kim2005a,Fu2007,Berg1979} and technological applications, such as swimming microrobots or microflow sensors \cite{Peyer2013,Attia2009,Schamel2014,Tottori2012,Zhang2009c,Fischer2011}.  Nature has perhaps best demonstrated the importance of small scale helix-fluid interactions through the evolution of helically shaped flagella, which are exploited by swimming microorganisms to move through their surrounding fluids \cite{Turner2000,Fujii2008,Armitage1999}.  At these length scales, structures function at low Reynolds number (i.e. inertia is negligible and viscous forces play a dominant role), \edit{defined as $Re=\rho v l / \eta$, where $\rho$ and $\eta$ are the fluid density and viscosity, respectively, $v$ is the flow velocity, and $l$ is a characteristic length scale. In these instances,} the helical structure is key to locomotive functionality; however many questions remain with regard to the fluid-helix interactions at these small length scales.

\begin{figure*}
\centering
\includegraphics[width=0.98\textwidth]{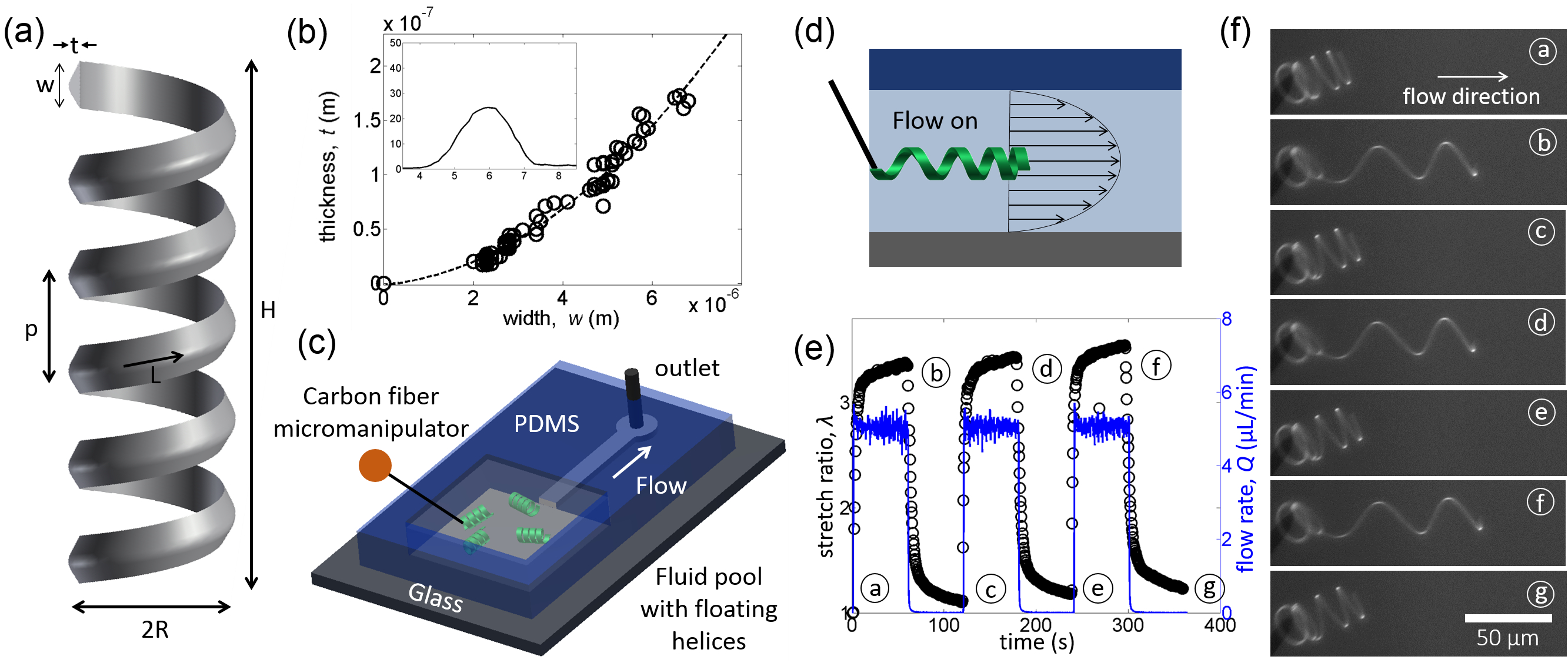}
\caption{(a) Geometry of a helical ribbon. \rev{(b) The relationship between the thickness and width of PMMA ribbons is best fit to a quadratic: $t=aw^{2}$ where $a=3530$ m$^{-1}$. Inset: 3D profile of ribbon cross-section measured by optical profilometry. y-axis is in nm and x-axis is in $\mu$m. Note the nanoscale thickness and microscale width.} (c) Experimental setup that allows helical ribbons to form in a large pool and be placed into a connected microfluidic channel and (d) the placement of the helix in the vertical center of the channel (at $v_{max}$).  (e) Measured stretch ratio and flow rate in a 3 cycle experiment with corresponding fluorescent images in (f). \label{fig1}}
\end{figure*}
While helices in low Reynolds number flows have been considered in several studies over the past couple decades, experimental work has focused mainly on macroscopic, non-deformable helical models in high viscosity fluids \cite{Liu2011,Kim2003,Rodenborn2013}, likely due to the difficulties in fabricating and analyzing microscopic systems in a controlled manner.  A natural bacterial flagellar filament is on the order of tens of nanometers in diameter and several microns long with bending stiffness in the range of $B \sim 10^{-24}$ to $10^{-21}$ N~m$^2$ \cite{Hoshikawa1985,Darnton2007,Trachtenberg1992}, values that have been measured through optical tweezer or crude flow experiments.  This low flexural stiffness results in drastic changes in shape of bacterial flagella observed experimentally under the motion of fluids \cite{Hoshikawa1985,Armitage1999,Turner2000,Coombs2002,Hotani1982}.  Moreover, the frictional coefficient that defines the relative resistance of motion between the solid and fluid is an important physical parameter for small helices in flow \cite{Fu2007,Kim2005a,Lauga2007,Lauga2009,Lighthill1976,Wada2009}.  Therefore, a microscopic experimental model that examines flexible helices in low Reynolds number flow, with the ability to predict and extract helical shape changes and frictional properties, would be exceedingly beneficial.

In this Communication, we examine the deformation of synthetically fabricated helical ribbons in controlled viscous flow with length scales and mechanical properties \edit{that approach those found in bacterial} flagella and microscale robots \cite{Turner2000,Fujii2008} (\textit{i.e.} microscale radius, nanoscale thickness). \edit{Here, taking the helical ribbon thickness $\sim$50~nm as the characteristic length scale for axial flow experiments, our approach allows for experiments in low Reynolds number even with strong flows ($\sim$10 mm/s), $Re \sim 10^{-4}$.} We discuss our findings in the framework of resistive force theory \cite{Gray1955,Lighthill1976} and demonstrate that, as expected, the size, shape and bending stiffness of a helical ribbon defines the axial deformation of microhelices in flow \cite{Kim2005a,Lighthill1976}. We quantify the non-uniform shape of a flexible helix deformed by viscous drag, showing that the pitch distribution transitions from linear to non-linear behavior within the same helix as a function of flow velocity. Our measurements allow us to assess validity of the resistive force theory and extract the effective frictional coefficient for microscale, flexible helices.

Consider a helical ribbon defined by its axial length ($H$), contour length ($L$), pitch ($p$), and radius ($R$), as well as its cross-section, which is defined by the ribbon width ($w$) and thickness ($t$), as illustrated in Fig. \ref{fig1}a. To create such structures, we recently reported a method that relies on spontaneous formation of helices from initially flat ribbons, driven by 2-phase elastocapillary deformation \cite{Pham2013}.  The ribbons are taken to be inextensible (\textit{i.e.} a fixed contour length), and under the condition that $t/w \ll 1$ and $w/L \ll 1$, helices form by bending in the direction of the nanoscale thickness (Figs. \ref{fig1}a \rev{and \ref{fig1}b}).  A key point to emphasize is that the preferred helical radius has a strong dependence on the ribbon thickness \cite{Pham2013}; hence the bending stiffness, $B=EI$ and the helix radius, $R$ are not independently controlled ($E$ being the Young's modulus and $I \sim wt^{3}$ being the second moment of area). This approach is advantageous since it provides versatility in controlling the helix geometry through control of fabrication parameters.

In our experiments, a flow rate ($Q$) is applied to a helix that is held in a microfluidic channel.  To fabricate the helices, ribbons are first prepared on a flat substrate by an evaporative assembly method \cite{Lee2013}.  We use a common glassy polymer as a model material: poly(methyl methacrylate) (PMMA, 120k g/mol) with fluorescent dye \edit{(Coumarin 153)} for imaging.  The ribbons are released into a pool of water, at which point they spontaneously form helices through a balance of surface tension and elasticity of the asymmetric cross-sectional geometry.  Details on helix formation and fabrication can be found in prior publications \cite{Pham2013,Lee2013,Kim2010a}.  A micromanipulator \edit{with a carbon fiber attached at its end} is subsequently used to fix one end of a helix and position it inside a microchannel (Fig. \ref{fig1}c) at the vertical center (\textit{i.e.} the center of the Poiseuille flow), which is 600 $\mu$m wide and 100 $\mu$m tall (Fig. \ref{fig1}d). \edit{The carbon fiber is brought into contact with the helix and is fixed due to non-specific interactions.}  The flow velocity ($v$) is taken to be $v_{max}$, where $v_{max}=3 v_{avg}/2$ and $v_{avg} = Q/A$, where $A$ is the channel cross-sectional area.

The designed setup has the advantage of measuring both flow rate and helix geometry simultaneously in real-time with a flow sensor and a fluorescence microscope.  In Figs. \ref{fig1}e and \ref{fig1}f, we present a typical flow cycle experiment to demonstrate helix shape recovery and flow control.  When the flow is turned on, the helix deforms along its helical axis in the direction of applied flow and in the absence of flow, the helix returns nearly to its original state.  This particular helix is cycled three times from 0 to 5 $\mu$L/min \rev{(corresponding to $v \sim 2$ mm/s)} and the flow rate history and the stretch ratio, $\lambda = H/H_{0}$, are plotted along with corresponding micrographs. Here, $H_{0}$ is the axial length of the helix in the absence of flow.  At point b, $\lambda \approx 3.4$ and recovers to point~c where $\lambda \approx 1.1$ when the flow is turned off for 90~s.  On the second cycle (point d), $\lambda \approx 3.4$ and recovers to $\lambda \approx 1.15$ and responds similarly in the third cycle, showing reversibility in our helices.  The small, irreversible deformations observed are likely associated with creep deformations within the ribbon material, but as shown below, these slight changes can be considered negligible for the focus of this work.

Deformation of a helix in an external flow is caused by the hydrodynamic drag forces acting at each point along its contour length. Following resistive force theory, the drag force per unit length is given by: $\textbf{f}=-\zeta_{\perp}\left [ \textbf{v}-\left (\textbf{t} \cdot \textbf{v}  \right ) \textbf{t} \right ] -\zeta_{\parallel} \left (\textbf{t} \cdot \textbf{v}  \right ) \textbf{t}$, where $\textbf{t}$ is the local tangent of the ribbon backbone, $\textbf{v}$ is the velocity of the fluid relative to the ribbon, and $\zeta_{\perp}$ and $\zeta_{\parallel}$ are the frictional coefficients that define the resistance to motion of the surrounding fluid in the normal and tangential directions from the ribbon, respectively \cite{Wada2009,Kim2005a}. These frictional coefficients are proportional to the viscosity ($\eta$) and a logarithmic correction dependent on the helical geometry~\cite{Lighthill1976}. In general, for very elongated objects, the ratio $\zeta_{\perp}/\zeta_{\parallel} \approx 2$. For the case of axial extension under flow, Kim and Powers \cite{Kim2005a} give an expression for the helix extension in the limit that $R/L$ is small and $\zeta_{\perp}/\zeta_{\parallel} \approx 2$:
\begin{equation}
\frac{\Delta H}{L} = \frac{\zeta_{\parallel} vR^{2}L}{B}
\label{powers}
\end{equation}
\noindent where the velocity $v$ is in the direction of the helical axis and the axial extension is defined as $\Delta H=H-H_{0}$.

\begin{figure}
\centering
\includegraphics[width=0.48\textwidth]{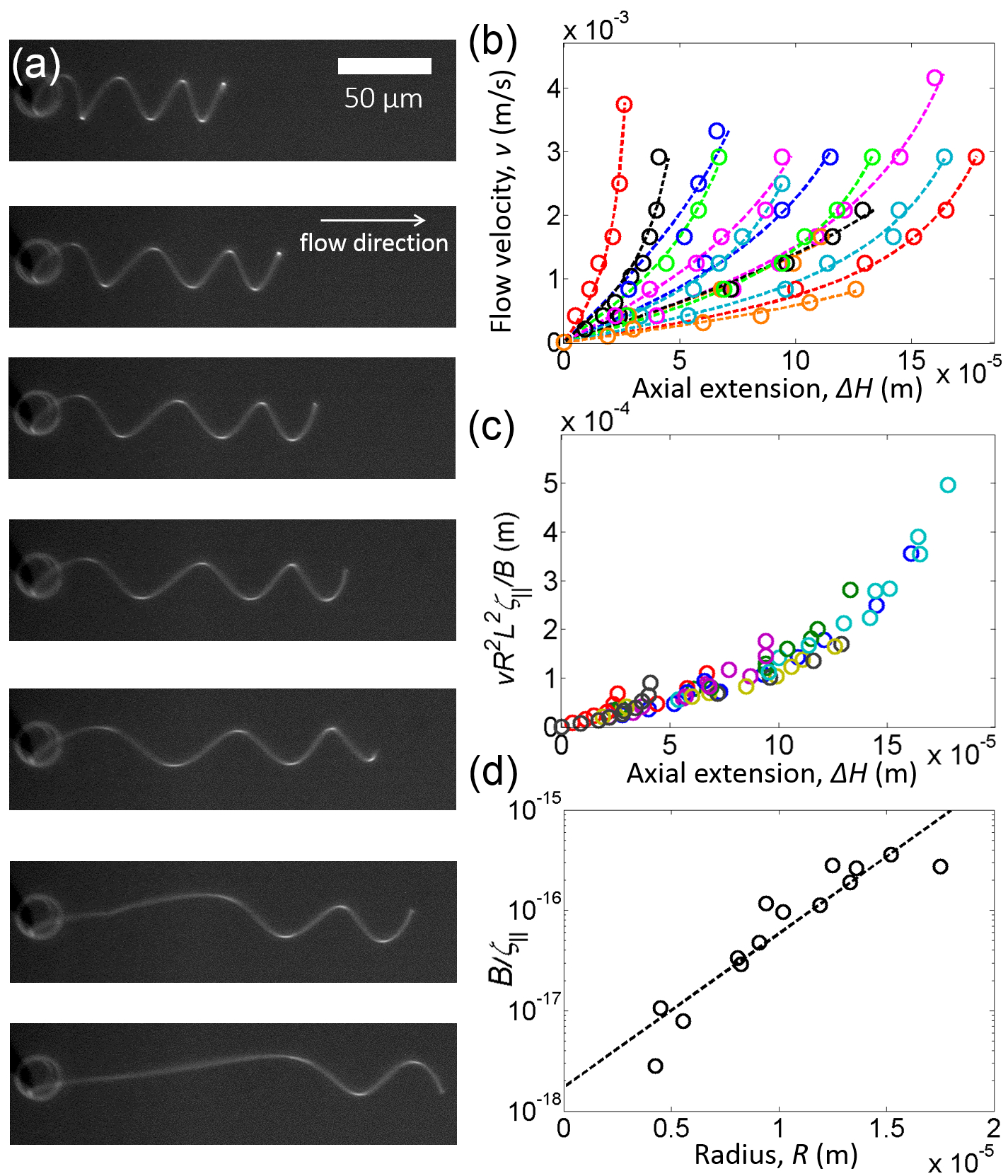}
\caption{(a) Fluorescent images of a helix with increasing flow velocities.  At higher velocity, the helix begins to lose turns by rotating its free end.  (b) Flow-extension curves for several helices, showing different extension due to the different helical dimensions (\textit{i.e.} $R$ and $L$). The second magenta data corresponds to (a). Dashed lines are a fit for a helix with non-linear, finite extensibility given in Eq. \ref{NFE}. \edit{(c) Data from (b) scaled by the helical geometries and determined $B/\zeta_{\parallel}$.} (d) Semilog plot of $B/\zeta_{\parallel}$ determined by the flow experiments as a function of the helix radius $R$. \label{fig2}}
\end{figure}

Guided by Eq. \ref{powers}, we measure $\Delta H$ as a function of~$v$. From Fig. \ref{fig2}b, the axial extension of the helices is non-linear with increasing flow velocity. We describe this non-linearity phenomenologically with non-linear finite extensibility \cite{WarnerJr.1972,Kroger2004} which leads to:
\begin{equation}
v=\frac{B}{\zeta_{\parallel} R^{2}L^{2}} \frac{\Delta H}{1-\left (\frac{\Delta H}{\Delta H_{max}}  \right )^{2}}
\label{NFE}
\end{equation}
\noindent where the maximum extension is taken to be $\Delta H_{max}=L-H_{0}$.  A typical experiment is shown in Fig. \ref{fig2}a.  Helices with a range of sizes were created to examine the effects of helix and ribbon geometry; these range between $R \approx 3-15 \mu$m, $p \approx 4-20 \mu$m, and $L \approx 55-420 \mu$m. $R$ is controlled by the bending stiffness $B=EI$ \cite{Pham2013}, where $E=2$~GPa is a typical value for PMMA \cite{Zeng2004}.  $R$ and $p$ are measured directly from the microscope images at zero flow rate and $L$ is determined by the helical relationship $L=N \sqrt{4 \pi^{2}R^{2}+p^{2}}$, where $N$ is the number of turns.  As expected, the assortment of helices display different flow-extension curves due to their varying shape and size, demonstrated in Fig. \ref{fig2}b.  The dashed lines represent the fit of Eq.~\ref{NFE} and our data are well fit to this relation.  Using the corresponding $R$ and $L$ values at zero flow rate, Eq.~\ref{NFE} leads to a best fit for $B/\zeta_{\parallel}$ for all helices. \edit{Scaling with the measured helical geometries and determined $B/\zeta_{\parallel}$ leads to a collapse of the data (Fig. \ref{fig2}c), validating the use of Eq.~\ref{NFE}.}  We determine the relationship between $B/\zeta_{\parallel}$ and $R$ for flow experiments by fitting to the expression $B/\zeta_{\parallel}=C\exp{(\alpha R)}$, giving $C=(1.7 \pm 0.7)$x$10^{-18}$~m$^{4}$~s$^{-1}$ and $\alpha=(3.5 \pm 0.4)$x$10^{5}$~m$^{-1}$ (Fig. \ref{fig2}d).

\begin{figure}
\centering
\includegraphics[width=0.49\textwidth]{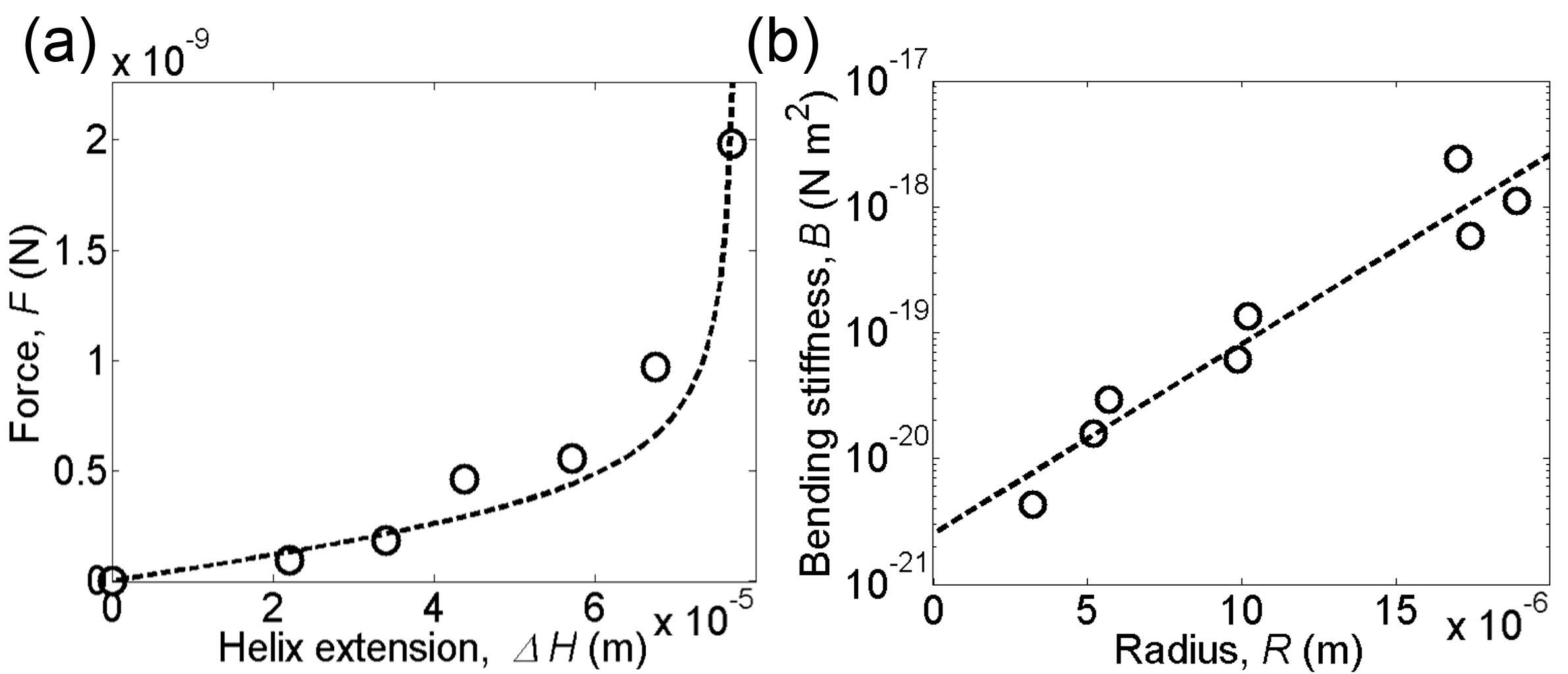}
\caption{\rev{(a) Force-extension curve for an end-loaded helix. (b) The determined relationship between $B$ and $R$ in end-loaded experiments.} \label{figSI}}
\end{figure}

To determine frictional coefficients, we quantify $B$ independently with a recently developed micromechanical tool to measure the end-loaded force-extension relationship \rev{(Fig. \ref{figSI}a)} of our helices for different helix geometries \rev{(see refs. \cite{Pham2013,Pham2014} for experimental details)}. Under end-loading conditions in the linear limit, the helical extension is given by \cite{Kim2005a}:
\begin{equation}
\frac{\Delta H}{L} = \frac{F R^{2}}{B}
\label{endloading}
\end{equation}
At high extension, the force-extension relationship is non-linear and follows expressions developed previously by Pham et al. \cite{Pham2014}\footnote{The expression of non-linearity used for this end-loading experiment is slightly different from the expression used for the flow experiments as the boundary conditions are different. $F= \frac{4\pi^{2}N^{2}BH}{L^{3}} \left [ \frac{\sqrt{1-\left ( H_{0}/L \right )^{2}}}{\sqrt{1-\left ( H/L \right )^{2}}} +M\right ]$ where $F$ is the force and the constant $M=2/(1+\nu)-1$ (where $\nu \approx 0.3$ is the Poisson's ratio). \cite{Pham2014} In the small strain limit where $H \ll L$, the force scales as $F\sim N^{2}BH/L^{3}$.  A geometric relationship for a helical structure holds that $R\sim L/N$, leading to $F \sim BH/R^{2}L$ used to determine $B$, identical to Eq. \ref{endloading}.}. A plot of $B$ vs. $R$ provides the empirical relation $B=B_{0}\exp{(\beta R)}$ with $B_{0}=(2.6 \pm 1.1)$x$10^{-21}$~N~m$^{2}$, \edit{which is comparable to bacterial flagella,} and $\beta=(3.5 \pm 0.5)$x$10^5$~m$^{-1}$ \rev{(Fig. \ref{figSI}b)}.  Importantly, we find $\alpha \approx \beta$, demonstrating that $\zeta_{\parallel}$ is independent of $R$ within our experimental resolution and parameter range.  Accordingly, a frictional drag coefficient can be quantitatively determined as $\zeta_{\parallel}=B_{0}/C=1.5 \pm 0.6$~mPa.s.

While different expressions of $\zeta_{\parallel}$ have been proposed by different researchers \cite{Wada2009,Kim2005a,Lauga2009,Lighthill1976}, the general relevant form for a circular cross-section is given by \cite{Lighthill1976}:
\begin{equation} \zeta_{\parallel}=\frac{2 \pi \eta}{\ln(2q/a)} \label{zetapar} \end{equation}
\noindent where $q$ is usually taken as 0.09$p$ and $a$ as the radius of the cylindrical fiber itself.  Since our helices' cross-sections are not circular, but rather a shallow triangular ribbon \rev{(Fig. \ref{fig1}b)}, we took the ribbon thickness to be the relevant length scale~$a$.  Although average ribbon dimensions are measured before transformation into helices, determining the nanoscale cross-sectional thickness of specific ribbons in their helical form is not possible with the optical microscope used to record the helix deformations.  Thus, the thickness $t$ is determined for specific helices by relating $R$, measured optically, to established relations for $B$ and measured aspect ratios of $t/w$ by AFM and optical profilometry \rev{(Fig. \ref{fig1}b)}. Taking these values for $t$ and the typical viscosity of water $\eta=1$ mPa.s, we calculate a theoretical $\zeta_{\parallel}$ for each helix using Eq.~\ref{zetapar}, providing $\zeta_{\parallel}=1.6-2.5$~mPa.s. This range is in reasonable agreement with our experimental results: $\zeta_{\parallel}=1.5 \pm 0.6$~mPa.s. It must be noted that since the frictional coefficient $\zeta_{\parallel}$ depends on the geometry of the helix as a logarithmic correction, the range of pitch and radius studied here are unfortunately not sufficient to resolve differences within our experimental resolution. 
\begin{figure}
\centering
\includegraphics[width=0.35\textwidth]{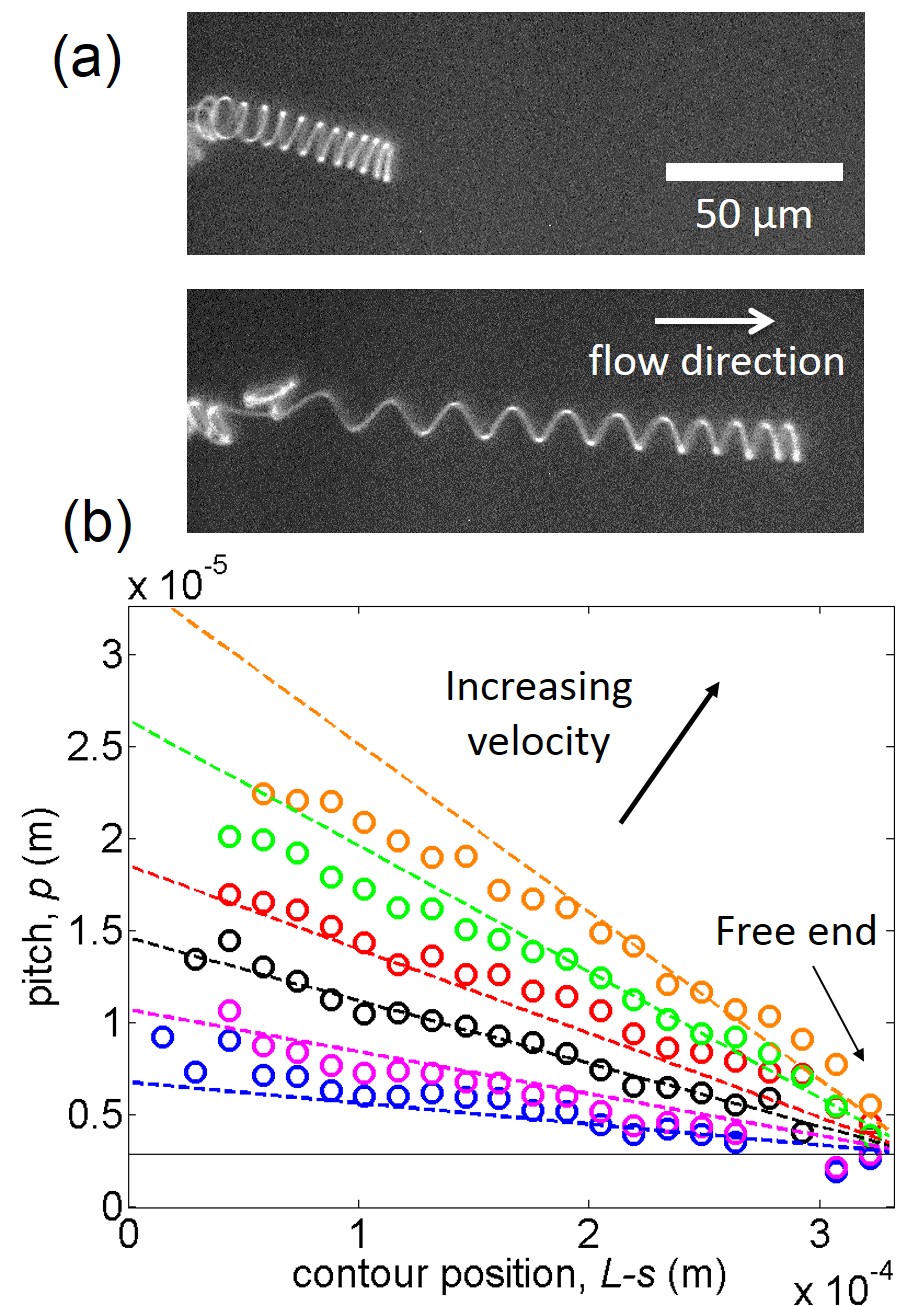}
\caption{(a) Helical ribbon with $R \approx$ 4.5 $\mu$m and $L \approx$ 320 $\mu$m in the absence of flow (top) and at $v=0.625$ mm/s (bottom). (b) Pitch as a function of position on the contour length corresponding to the helix in (a) for different flow velocity. The dotted lines are calculated from Eq. \ref{pitch} and the thin black represents the calculated $p_0$. The error on pitch measurements is within the size of the points on the graph ($<1$ $\mu$m). \label{fig3}}
\end{figure}

Aside from their global extension, flexible helices display non-uniform shape distributions when deformed in fluid flow, which has also been observed in helical flagella \cite{Armitage1999,Hoshikawa1985}.  More specifically,  it is observed that the turns are most stretched at the fixed end and continuously become less stretched along the helix approaching the free end.  This is clearly visualized in an experiment of a long helix with several turns as shown in Fig.~\ref{fig3}, where $L \approx$ 320 $\mu$m and $R \approx$ 4.5 $\mu$m, and at an applied flow velocity of $v=0.625$~mm/s. Such non-uniform shapes are readily explained by distributed loading of helices: under flow, the force applied to a small element of a helix consists of local hydrodynamic drag on the element and the force accumulating along the helix from the free end. Mechanical equilibrium is then ensured by the equal and opposite force applied to the element by the rest of the helix that is further away from the free end. If we assume that the local hydrodynamic drag is independent of the position along the helix, as in the resistive-force theory discussed above, the total force applied to an element of the helix from the free end grows linearly with the contour length~$s$, measured from the free end. Consequently, the local pitch of the helix also grows linearly with $s$, as shown below. This situation is analogous to stretching of low-stiffness springs under gravity \cite{Cross2012,Miller2014}.  \edit{Similarly, variations in the radius are observed when the local stretch is sufficiently high, as demonstrated near the fixed end of the helix in Fig. \ref{fig3}.}

To quantitatively examine the shape distribution of our helices, we measure the local pitch~$p(s)$ by calculating the distance between the outmost points of neighboring turns along the helix. At zero flow rate, the pitch~$p_0$ is constant along the helix within small experimental variations. Under flow, $p(s)$ can be estimated as the difference of the axial displacement of the points $s+l/2$ and $s-l/2$ obtained from Eq.~\ref{powers}:
\begin{eqnarray}
&&p(s) = \Delta H(s+l/2) - \Delta H(s-l/2) +p_0\nonumber \\
&& \qquad = \frac{ 2\zeta_{\parallel} v R^{2}}{B} l s + p_0,
\label{pitch}
\end{eqnarray}
where $l$ is the contour length of one pitch, which we assume to be constant. This assumption is justified as long as no turns are lost during the experiment and irreversible deformations are negligible.  In Fig.~\ref{fig3}b, we plot Eq.~\ref{pitch} with the corresponding values of $\zeta_{\parallel}$, $R$ and $B$ determined by our flow experiment for the different flow velocities. In  Fig. \ref{fig3}b, one can observe that for small velocities, the pitch vs. position dependence is well described within the experimental errors by Eq.~\ref{pitch}.  Consequently, the approximation of linear axial extension (Eq.~\ref{powers}) holds at low velocities; however, deviations are observed at high flow velocities. Here the pitch distribution in the helix section experiencing the highest forces (closer to the fixed end) deviates from the linear approximation of Eq.~\ref{pitch}. In this part of the helix, measured pitches are smaller than predicted, corresponding to geometric strain stiffening at large extension also seen in Fig.~\ref{fig2}b. Notably, our results demonstrate a spatial manifestation of the crossover between linear and non-linear behavior of a helix under distributed loading.

\begin{figure}
\centering
\includegraphics[width=0.35\textwidth]{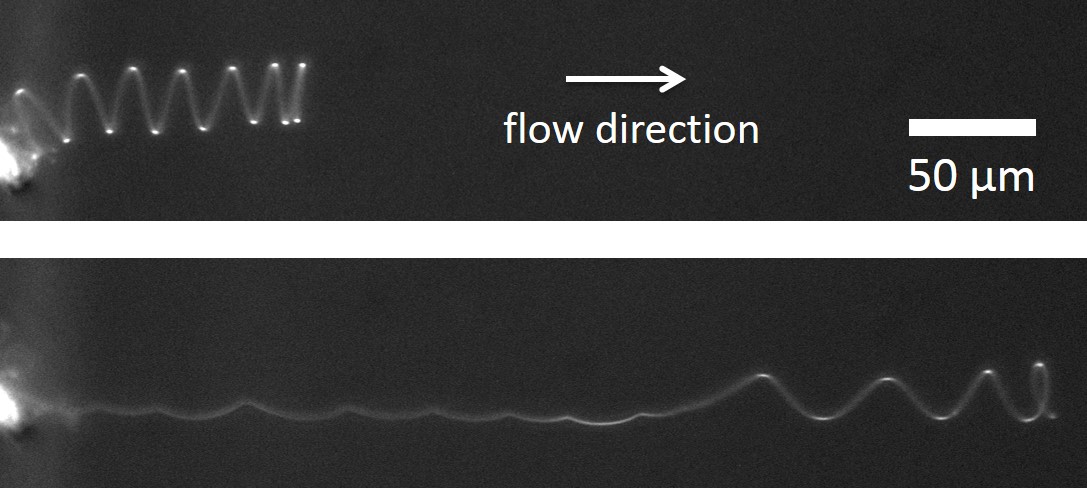}
\caption{Qualitative example of a prominent helical transition from a global helical geometry to local uncoiled and coiled configurations at high velocity (of order $\sim 10$ mm/s). \label{fig4}}
\end{figure}

Finally, we interestingly observe large shape changes in helices at high velocities.  Under these stronger forces, the helical shape uncoils to lose turns near the point of attachment (Fig. \ref{fig2}a).  More evident helical instabilities are observed through localized transitions of coiled to uncoiled helical geometry, as shown in Fig. \ref{fig4}.  These drastic deformations are reversible; when the flow is turned off, the helix relaxes to a shape that again is nearly identical to the initial helix.  Similar transitions of helical and straightened geometries have been observed in torque-free, end-loaded experiments on cholesterol helical ribbons \cite{Smith2001a} as well as in rods of preferred curvature under gravity \cite{Miller2014}. \edit{Moreover, similar conformational transitions at high flow rates are observed with flexible polymer chains \cite{Brochard-Wyart1996}.} Thus, future studies will focus on these helical transitions in viscous flow, which may provide important insight into unstable transitions that exist in helical systems found in nature.

Overall, we have introduced a microscopic model system to measure the deformation, shape and frictional properties of flexible helices in low Reynolds number viscous flow and find that the global axial deformation is consistent with existing theory \cite{Kim2005a}.  We demonstrate that with known ribbon properties and helical configurations, the shape distribution can be quantitatively predicted.  Moreover, our experimental platform presents opportunities for theoretical advances on flexible helices in low Reynolds number flow; in particular, the effects of fluid viscosity or viscoelasticity, the friction and flow around deformable helices, the global-to-local helical shape transitions, and potentially the effects of cross-sectional geometry. Understanding these general helical behaviors both experimentally and theoretically will lead to fundamental insights on natural helices, like flagella, as well as the development of synthetic helices, like swimming microbots in fluid environments. Well-characterized helices can also be used to measure local forces in flows of simple or complex fluids where local velocities can readily be measured by various techniques, like PIV or particle-tracking, while measuring local stresses presents a significant technological challenge~\cite{Liu2010b}.

This work was supported by the Army Research Office (W911NF-14-1-0185) and a Chateaubriand Fellowship granted by the French Embassy in the United States of America. A.M. acknowledges support from the UK Engineering and Physical Sciences Research Council (EP/I004262/1). The authors thank J. Heuvingh and G.M. Grason for thoughtful discussion and J. Gachelin for providing photolithography molds for microchannel fabrication.

\bibliography{Pham_References_Helices}

\begin{thebibliography}{40}%
\makeatletter
\providecommand \@ifxundefined [1]{%
 \@ifx{#1\undefined}
}%
\providecommand \@ifnum [1]{%
 \ifnum #1\expandafter \@firstoftwo
 \else \expandafter \@secondoftwo
 \fi
}%
\providecommand \@ifx [1]{%
 \ifx #1\expandafter \@firstoftwo
 \else \expandafter \@secondoftwo
 \fi
}%
\providecommand \natexlab [1]{#1}%
\providecommand \enquote  [1]{``#1''}%
\providecommand \bibnamefont  [1]{#1}%
\providecommand \bibfnamefont [1]{#1}%
\providecommand \citenamefont [1]{#1}%
\providecommand \href@noop [0]{\@secondoftwo}%
\providecommand \href [0]{\begingroup \@sanitize@url \@href}%
\providecommand \@href[1]{\@@startlink{#1}\@@href}%
\providecommand \@@href[1]{\endgroup#1\@@endlink}%
\providecommand \@sanitize@url [0]{\catcode `\\12\catcode `\$12\catcode
  `\&12\catcode `\#12\catcode `\^12\catcode `\_12\catcode `\%12\relax}%
\providecommand \@@startlink[1]{}%
\providecommand \@@endlink[0]{}%
\providecommand \url  [0]{\begingroup\@sanitize@url \@url }%
\providecommand \@url [1]{\endgroup\@href {#1}{\urlprefix }}%
\providecommand \urlprefix  [0]{URL }%
\providecommand \Eprint [0]{\href }%
\providecommand \doibase [0]{http://dx.doi.org/}%
\providecommand \selectlanguage [0]{\@gobble}%
\providecommand \bibinfo  [0]{\@secondoftwo}%
\providecommand \bibfield  [0]{\@secondoftwo}%
\providecommand \translation [1]{[#1]}%
\providecommand \BibitemOpen [0]{}%
\providecommand \bibitemStop [0]{}%
\providecommand \bibitemNoStop [0]{.\EOS\space}%
\providecommand \EOS [0]{\spacefactor3000\relax}%
\providecommand \BibitemShut  [1]{\csname bibitem#1\endcsname}%
\let\auto@bib@innerbib\@empty
\bibitem [{\citenamefont {Darwin}(1875)}]{Darwin1875}%
  \BibitemOpen
  \bibfield  {author} {\bibinfo {author} {\bibfnamefont {C.}~\bibnamefont
  {Darwin}},\ }\href@noop {} {\emph {\bibinfo {title} {{The movements and
  habits of climbing plants}}}},\ \bibinfo {edition} {2nd}\ ed.\ (\bibinfo
  {publisher} {London: John Murray},\ \bibinfo {year} {1875})\BibitemShut
  {NoStop}%
\bibitem [{\citenamefont {Wada}\ and\ \citenamefont {Netz}(2009)}]{Wada2009}%
  \BibitemOpen
  \bibfield  {author} {\bibinfo {author} {\bibfnamefont {H.}~\bibnamefont
  {Wada}}\ and\ \bibinfo {author} {\bibfnamefont {R.}~\bibnamefont {Netz}},\
  }\href {\doibase 10.1103/PhysRevE.80.021921} {\bibfield  {journal} {\bibinfo
  {journal} {Physical Review E}\ }\textbf {\bibinfo {volume} {80}},\ \bibinfo
  {pages} {021921} (\bibinfo {year} {2009})}\BibitemShut {NoStop}%
\bibitem [{\citenamefont {Teran}\ \emph {et~al.}(2010)\citenamefont {Teran},
  \citenamefont {Fauci},\ and\ \citenamefont {Shelley}}]{Teran2010}%
  \BibitemOpen
  \bibfield  {author} {\bibinfo {author} {\bibfnamefont {J.}~\bibnamefont
  {Teran}}, \bibinfo {author} {\bibfnamefont {L.}~\bibnamefont {Fauci}}, \ and\
  \bibinfo {author} {\bibfnamefont {M.}~\bibnamefont {Shelley}},\ }\href
  {\doibase 10.1103/PhysRevLett.104.038101} {\bibfield  {journal} {\bibinfo
  {journal} {Physical Review Letters}\ }\textbf {\bibinfo {volume} {104}},\
  \bibinfo {pages} {038101} (\bibinfo {year} {2010})}\BibitemShut {NoStop}%
\bibitem [{\citenamefont {Rodenborn}\ \emph {et~al.}(2013)\citenamefont
  {Rodenborn}, \citenamefont {Chen}, \citenamefont {Swinney},\ and\
  \citenamefont {Zhang}}]{Rodenborn2013}%
  \BibitemOpen
  \bibfield  {author} {\bibinfo {author} {\bibfnamefont {B.}~\bibnamefont
  {Rodenborn}}, \bibinfo {author} {\bibfnamefont {C.-h.}\ \bibnamefont {Chen}},
  \bibinfo {author} {\bibfnamefont {H.}~\bibnamefont {Swinney}}, \ and\
  \bibinfo {author} {\bibfnamefont {H.}~\bibnamefont {Zhang}},\ }\href
  {\doibase
  10.1073/pnas.1219831110/-/DCSupplemental.www.pnas.org/cgi/doi/10.1073/pnas.1219831110}
  {\bibfield  {journal} {\bibinfo  {journal} {Proceedings of the National
  Academy of Sciences of the United States of America}\ }\textbf {\bibinfo
  {volume} {110}},\ \bibinfo {pages} {338} (\bibinfo {year}
  {2013})}\BibitemShut {NoStop}%
\bibitem [{\citenamefont {Liu}\ \emph {et~al.}(2011)\citenamefont {Liu},
  \citenamefont {Powers},\ and\ \citenamefont {Breuer}}]{Liu2011}%
  \BibitemOpen
  \bibfield  {author} {\bibinfo {author} {\bibfnamefont {B.}~\bibnamefont
  {Liu}}, \bibinfo {author} {\bibfnamefont {T.~R.}\ \bibnamefont {Powers}}, \
  and\ \bibinfo {author} {\bibfnamefont {K.~S.}\ \bibnamefont {Breuer}},\
  }\href {\doibase 10.1073/pnas.1113082108} {\bibfield  {journal} {\bibinfo
  {journal} {Proceedings of the National Academy of Sciences}\ }\textbf
  {\bibinfo {volume} {108}},\ \bibinfo {pages} {19516} (\bibinfo {year}
  {2011})}\BibitemShut {NoStop}%
\bibitem [{\citenamefont {Lauga}\ and\ \citenamefont
  {Powers}(2009)}]{Lauga2009}%
  \BibitemOpen
  \bibfield  {author} {\bibinfo {author} {\bibfnamefont {E.}~\bibnamefont
  {Lauga}}\ and\ \bibinfo {author} {\bibfnamefont {T.~R.}\ \bibnamefont
  {Powers}},\ }\href {\doibase 10.1088/0034-4885/72/9/096601} {\bibfield
  {journal} {\bibinfo  {journal} {Reports on Progress in Physics}\ }\textbf
  {\bibinfo {volume} {72}},\ \bibinfo {pages} {096601} (\bibinfo {year}
  {2009})}\BibitemShut {NoStop}%
\bibitem [{\citenamefont {Kim}\ and\ \citenamefont {Powers}(2005)}]{Kim2005a}%
  \BibitemOpen
  \bibfield  {author} {\bibinfo {author} {\bibfnamefont {M.}~\bibnamefont
  {Kim}}\ and\ \bibinfo {author} {\bibfnamefont {T.}~\bibnamefont {Powers}},\
  }\href {\doibase 10.1103/PhysRevE.71.021914} {\bibfield  {journal} {\bibinfo
  {journal} {Physical Review E}\ }\textbf {\bibinfo {volume} {71}},\ \bibinfo
  {pages} {021914} (\bibinfo {year} {2005})}\BibitemShut {NoStop}%
\bibitem [{\citenamefont {Fu}\ \emph {et~al.}(2007)\citenamefont {Fu},
  \citenamefont {Powers},\ and\ \citenamefont {Wolgemuth}}]{Fu2007}%
  \BibitemOpen
  \bibfield  {author} {\bibinfo {author} {\bibfnamefont {H.}~\bibnamefont
  {Fu}}, \bibinfo {author} {\bibfnamefont {T.}~\bibnamefont {Powers}}, \ and\
  \bibinfo {author} {\bibfnamefont {C.}~\bibnamefont {Wolgemuth}},\ }\href
  {\doibase 10.1103/PhysRevLett.99.258101} {\bibfield  {journal} {\bibinfo
  {journal} {Physical Review Letters}\ }\textbf {\bibinfo {volume} {99}},\
  \bibinfo {pages} {258101} (\bibinfo {year} {2007})}\BibitemShut {NoStop}%
\bibitem [{\citenamefont {Berg}\ and\ \citenamefont {Turner}(1979)}]{Berg1979}%
  \BibitemOpen
  \bibfield  {author} {\bibinfo {author} {\bibfnamefont {H.}~\bibnamefont
  {Berg}}\ and\ \bibinfo {author} {\bibfnamefont {L.}~\bibnamefont {Turner}},\
  }\href {http://www.nature.com/nature/journal/v278/n5702/abs/278349a0.html}
  {\bibfield  {journal} {\bibinfo  {journal} {Nature}\ }\textbf {\bibinfo
  {volume} {278}},\ \bibinfo {pages} {349} (\bibinfo {year}
  {1979})}\BibitemShut {NoStop}%
\bibitem [{\citenamefont {Peyer}\ \emph {et~al.}(2013)\citenamefont {Peyer},
  \citenamefont {Zhang},\ and\ \citenamefont {Nelson}}]{Peyer2013}%
  \BibitemOpen
  \bibfield  {author} {\bibinfo {author} {\bibfnamefont {K.~E.}\ \bibnamefont
  {Peyer}}, \bibinfo {author} {\bibfnamefont {L.}~\bibnamefont {Zhang}}, \ and\
  \bibinfo {author} {\bibfnamefont {B.~J.}\ \bibnamefont {Nelson}},\ }\href
  {\doibase 10.1039/c2nr32554c} {\bibfield  {journal} {\bibinfo  {journal}
  {Nanoscale}\ }\textbf {\bibinfo {volume} {5}},\ \bibinfo {pages} {1259}
  (\bibinfo {year} {2013})}\BibitemShut {NoStop}%
\bibitem [{\citenamefont {Attia}\ \emph {et~al.}(2009)\citenamefont {Attia},
  \citenamefont {Pregibon}, \citenamefont {Doyle}, \citenamefont {Viovy},\ and\
  \citenamefont {Bartolo}}]{Attia2009}%
  \BibitemOpen
  \bibfield  {author} {\bibinfo {author} {\bibfnamefont {R.}~\bibnamefont
  {Attia}}, \bibinfo {author} {\bibfnamefont {D.~C.}\ \bibnamefont {Pregibon}},
  \bibinfo {author} {\bibfnamefont {P.~S.}\ \bibnamefont {Doyle}}, \bibinfo
  {author} {\bibfnamefont {J.-L.}\ \bibnamefont {Viovy}}, \ and\ \bibinfo
  {author} {\bibfnamefont {D.}~\bibnamefont {Bartolo}},\ }\href@noop {}
  {\bibfield  {journal} {\bibinfo  {journal} {Lab on a Chip}\ }\textbf
  {\bibinfo {volume} {9}},\ \bibinfo {pages} {1213} (\bibinfo {year}
  {2009})}\BibitemShut {NoStop}%
\bibitem [{\citenamefont {Schamel}\ \emph {et~al.}(2014)\citenamefont
  {Schamel}, \citenamefont {Mark}, \citenamefont {Gibbs}, \citenamefont
  {Miksch}, \citenamefont {Morozov}, \citenamefont {Leshansky},\ and\
  \citenamefont {Fischer}}]{Schamel2014}%
  \BibitemOpen
  \bibfield  {author} {\bibinfo {author} {\bibfnamefont {D.}~\bibnamefont
  {Schamel}}, \bibinfo {author} {\bibfnamefont {A.~G.}\ \bibnamefont {Mark}},
  \bibinfo {author} {\bibfnamefont {J.~G.}\ \bibnamefont {Gibbs}}, \bibinfo
  {author} {\bibfnamefont {C.}~\bibnamefont {Miksch}}, \bibinfo {author}
  {\bibfnamefont {K.~I.}\ \bibnamefont {Morozov}}, \bibinfo {author}
  {\bibfnamefont {A.~M.}\ \bibnamefont {Leshansky}}, \ and\ \bibinfo {author}
  {\bibfnamefont {P.}~\bibnamefont {Fischer}},\ }\href {\doibase
  10.1021/nn502360t} {\bibfield  {journal} {\bibinfo  {journal} {ACS nano}\
  }\textbf {\bibinfo {volume} {8}},\ \bibinfo {pages} {8794} (\bibinfo {year}
  {2014})}\BibitemShut {NoStop}%
\bibitem [{\citenamefont {Tottori}\ \emph {et~al.}(2012)\citenamefont
  {Tottori}, \citenamefont {Zhang}, \citenamefont {Qiu}, \citenamefont
  {Krawczyk}, \citenamefont {Franco-Obreg\'{o}n},\ and\ \citenamefont
  {Nelson}}]{Tottori2012}%
  \BibitemOpen
  \bibfield  {author} {\bibinfo {author} {\bibfnamefont {S.}~\bibnamefont
  {Tottori}}, \bibinfo {author} {\bibfnamefont {L.}~\bibnamefont {Zhang}},
  \bibinfo {author} {\bibfnamefont {F.}~\bibnamefont {Qiu}}, \bibinfo {author}
  {\bibfnamefont {K.~K.}\ \bibnamefont {Krawczyk}}, \bibinfo {author}
  {\bibfnamefont {A.}~\bibnamefont {Franco-Obreg\'{o}n}}, \ and\ \bibinfo
  {author} {\bibfnamefont {B.~J.}\ \bibnamefont {Nelson}},\ }\href {\doibase
  10.1002/adma.201103818} {\bibfield  {journal} {\bibinfo  {journal} {Advanced
  Materials}\ }\textbf {\bibinfo {volume} {24}},\ \bibinfo {pages} {811}
  (\bibinfo {year} {2012})}\BibitemShut {NoStop}%
\bibitem [{\citenamefont {Zhang}\ \emph {et~al.}(2009)\citenamefont {Zhang},
  \citenamefont {Abbott}, \citenamefont {Dong}, \citenamefont {Peyer},
  \citenamefont {Kratochvil}, \citenamefont {Zhang}, \citenamefont {Bergeles},\
  and\ \citenamefont {Nelson}}]{Zhang2009c}%
  \BibitemOpen
  \bibfield  {author} {\bibinfo {author} {\bibfnamefont {L.}~\bibnamefont
  {Zhang}}, \bibinfo {author} {\bibfnamefont {J.~J.}\ \bibnamefont {Abbott}},
  \bibinfo {author} {\bibfnamefont {L.}~\bibnamefont {Dong}}, \bibinfo {author}
  {\bibfnamefont {K.~E.}\ \bibnamefont {Peyer}}, \bibinfo {author}
  {\bibfnamefont {B.~E.}\ \bibnamefont {Kratochvil}}, \bibinfo {author}
  {\bibfnamefont {H.}~\bibnamefont {Zhang}}, \bibinfo {author} {\bibfnamefont
  {C.}~\bibnamefont {Bergeles}}, \ and\ \bibinfo {author} {\bibfnamefont
  {B.~J.}\ \bibnamefont {Nelson}},\ }\href {\doibase 10.1021/nl901869j}
  {\bibfield  {journal} {\bibinfo  {journal} {Nano letters}\ }\textbf {\bibinfo
  {volume} {9}},\ \bibinfo {pages} {3663} (\bibinfo {year} {2009})}\BibitemShut
  {NoStop}%
\bibitem [{\citenamefont {Fischer}\ and\ \citenamefont
  {Ghosh}(2011)}]{Fischer2011}%
  \BibitemOpen
  \bibfield  {author} {\bibinfo {author} {\bibfnamefont {P.}~\bibnamefont
  {Fischer}}\ and\ \bibinfo {author} {\bibfnamefont {A.}~\bibnamefont
  {Ghosh}},\ }\href {\doibase 10.1039/c0nr00566e} {\bibfield  {journal}
  {\bibinfo  {journal} {Nanoscale}\ }\textbf {\bibinfo {volume} {3}},\ \bibinfo
  {pages} {557} (\bibinfo {year} {2011})}\BibitemShut {NoStop}%
\bibitem [{\citenamefont {Turner}\ \emph {et~al.}(2000)\citenamefont {Turner},
  \citenamefont {Ryu},\ and\ \citenamefont {Berg}}]{Turner2000}%
  \BibitemOpen
  \bibfield  {author} {\bibinfo {author} {\bibfnamefont {L.}~\bibnamefont
  {Turner}}, \bibinfo {author} {\bibfnamefont {W.}~\bibnamefont {Ryu}}, \ and\
  \bibinfo {author} {\bibfnamefont {H.}~\bibnamefont {Berg}},\ }\href {\doibase
  10.1128/JB.182.10.2793-2801.2000.Updated} {\bibfield  {journal} {\bibinfo
  {journal} {Journal of Bacteriology}\ }\textbf {\bibinfo {volume} {182}}
  (\bibinfo {year} {2000}),\
  10.1128/JB.182.10.2793-2801.2000.Updated}\BibitemShut {NoStop}%
\bibitem [{\citenamefont {Fujii}\ \emph {et~al.}(2008)\citenamefont {Fujii},
  \citenamefont {Shibata},\ and\ \citenamefont {Aizawa}}]{Fujii2008}%
  \BibitemOpen
  \bibfield  {author} {\bibinfo {author} {\bibfnamefont {M.}~\bibnamefont
  {Fujii}}, \bibinfo {author} {\bibfnamefont {S.}~\bibnamefont {Shibata}}, \
  and\ \bibinfo {author} {\bibfnamefont {S.-I.}\ \bibnamefont {Aizawa}},\
  }\href {\doibase 10.1016/j.jmb.2008.04.012} {\bibfield  {journal} {\bibinfo
  {journal} {Journal of Molecular Biology}\ }\textbf {\bibinfo {volume}
  {379}},\ \bibinfo {pages} {273} (\bibinfo {year} {2008})}\BibitemShut
  {NoStop}%
\bibitem [{\citenamefont {Armitage}\ \emph {et~al.}(1999)\citenamefont
  {Armitage}, \citenamefont {Pitta}, \citenamefont {Vigeant}, \citenamefont
  {Packer},\ and\ \citenamefont {Ford}}]{Armitage1999}%
  \BibitemOpen
  \bibfield  {author} {\bibinfo {author} {\bibfnamefont {J.~P.}\ \bibnamefont
  {Armitage}}, \bibinfo {author} {\bibfnamefont {T.~P.}\ \bibnamefont {Pitta}},
  \bibinfo {author} {\bibfnamefont {M.~a.}\ \bibnamefont {Vigeant}}, \bibinfo
  {author} {\bibfnamefont {H.~L.}\ \bibnamefont {Packer}}, \ and\ \bibinfo
  {author} {\bibfnamefont {R.~M.}\ \bibnamefont {Ford}},\ }\href
  {http://www.pubmedcentral.nih.gov/articlerender.fcgi?artid=93968\&tool=pmcentrez\&rendertype=abstract}
  {\bibfield  {journal} {\bibinfo  {journal} {Journal of Bacteriology}\
  }\textbf {\bibinfo {volume} {181}},\ \bibinfo {pages} {4825} (\bibinfo {year}
  {1999})}\BibitemShut {NoStop}%
\bibitem [{\citenamefont {Kim}\ \emph {et~al.}(2003)\citenamefont {Kim},
  \citenamefont {Bird}, \citenamefont {{Van Parys}}, \citenamefont {Breuer},\
  and\ \citenamefont {Powers}}]{Kim2003}%
  \BibitemOpen
  \bibfield  {author} {\bibinfo {author} {\bibfnamefont {M.}~\bibnamefont
  {Kim}}, \bibinfo {author} {\bibfnamefont {J.~C.}\ \bibnamefont {Bird}},
  \bibinfo {author} {\bibfnamefont {A.~J.}\ \bibnamefont {{Van Parys}}},
  \bibinfo {author} {\bibfnamefont {K.~S.}\ \bibnamefont {Breuer}}, \ and\
  \bibinfo {author} {\bibfnamefont {T.~R.}\ \bibnamefont {Powers}},\ }\href
  {\doibase 10.1073/pnas.2633596100} {\bibfield  {journal} {\bibinfo  {journal}
  {Proceedings of the National Academy of Sciences}\ }\textbf {\bibinfo
  {volume} {100}},\ \bibinfo {pages} {15481} (\bibinfo {year}
  {2003})}\BibitemShut {NoStop}%
\bibitem [{\citenamefont {Hoshikawa}\ and\ \citenamefont
  {Kamiya}(1985)}]{Hoshikawa1985}%
  \BibitemOpen
  \bibfield  {author} {\bibinfo {author} {\bibfnamefont {H.}~\bibnamefont
  {Hoshikawa}}\ and\ \bibinfo {author} {\bibfnamefont {R.}~\bibnamefont
  {Kamiya}},\ }\href
  {http://www.sciencedirect.com/science/article/pii/0301462285800387}
  {\bibfield  {journal} {\bibinfo  {journal} {Biophysical Chemistry}\ }\textbf
  {\bibinfo {volume} {22}},\ \bibinfo {pages} {159} (\bibinfo {year}
  {1985})}\BibitemShut {NoStop}%
\bibitem [{\citenamefont {Darnton}\ and\ \citenamefont
  {Berg}(2007)}]{Darnton2007}%
  \BibitemOpen
  \bibfield  {author} {\bibinfo {author} {\bibfnamefont {N.~C.}\ \bibnamefont
  {Darnton}}\ and\ \bibinfo {author} {\bibfnamefont {H.~C.}\ \bibnamefont
  {Berg}},\ }\href {\doibase 10.1529/biophysj.106.094037} {\bibfield  {journal}
  {\bibinfo  {journal} {Biophysical Journal}\ }\textbf {\bibinfo {volume}
  {92}},\ \bibinfo {pages} {2230} (\bibinfo {year} {2007})}\BibitemShut
  {NoStop}%
\bibitem [{\citenamefont {Trachtenberg}\ and\ \citenamefont
  {Hammel}(1992)}]{Trachtenberg1992}%
  \BibitemOpen
  \bibfield  {author} {\bibinfo {author} {\bibfnamefont {S.}~\bibnamefont
  {Trachtenberg}}\ and\ \bibinfo {author} {\bibfnamefont {I.}~\bibnamefont
  {Hammel}},\ }\href
  {http://www.sciencedirect.com/science/article/pii/104784779290063G}
  {\bibfield  {journal} {\bibinfo  {journal} {Journal of structural biology}\
  }\textbf {\bibinfo {volume} {109}},\ \bibinfo {pages} {18} (\bibinfo {year}
  {1992})}\BibitemShut {NoStop}%
\bibitem [{\citenamefont {Coombs}\ \emph {et~al.}(2002)\citenamefont {Coombs},
  \citenamefont {Huber}, \citenamefont {Kessler},\ and\ \citenamefont
  {Goldstein}}]{Coombs2002}%
  \BibitemOpen
  \bibfield  {author} {\bibinfo {author} {\bibfnamefont {D.}~\bibnamefont
  {Coombs}}, \bibinfo {author} {\bibfnamefont {G.}~\bibnamefont {Huber}},
  \bibinfo {author} {\bibfnamefont {J.~O.}\ \bibnamefont {Kessler}}, \ and\
  \bibinfo {author} {\bibfnamefont {R.~E.}\ \bibnamefont {Goldstein}},\ }\href
  {\doibase 10.1103/PhysRevLett.89.118102} {\bibfield  {journal} {\bibinfo
  {journal} {Physical review letters}\ }\textbf {\bibinfo {volume} {89}},\
  \bibinfo {pages} {118102} (\bibinfo {year} {2002})}\BibitemShut {NoStop}%
\bibitem [{\citenamefont {Hotani}(1982)}]{Hotani1982}%
  \BibitemOpen
  \bibfield  {author} {\bibinfo {author} {\bibfnamefont {H.}~\bibnamefont
  {Hotani}},\ }\href@noop {} {\bibfield  {journal} {\bibinfo  {journal}
  {Journal of Molecular Biology}\ }\textbf {\bibinfo {volume} {156}},\ \bibinfo
  {pages} {791} (\bibinfo {year} {1982})}\BibitemShut {NoStop}%
\bibitem [{\citenamefont {Lauga}(2007)}]{Lauga2007}%
  \BibitemOpen
  \bibfield  {author} {\bibinfo {author} {\bibfnamefont {E.}~\bibnamefont
  {Lauga}},\ }\href {\doibase 10.1103/PhysRevE.75.041916} {\bibfield  {journal}
  {\bibinfo  {journal} {Physical Review E}\ }\textbf {\bibinfo {volume} {75}},\
  \bibinfo {pages} {041916} (\bibinfo {year} {2007})}\BibitemShut {NoStop}%
\bibitem [{\citenamefont {Lighthill}(1976)}]{Lighthill1976}%
  \BibitemOpen
  \bibfield  {author} {\bibinfo {author} {\bibfnamefont {J.}~\bibnamefont
  {Lighthill}},\ }\href@noop {} {\bibfield  {journal} {\bibinfo  {journal}
  {SIAM Review}\ }\textbf {\bibinfo {volume} {18}},\ \bibinfo {pages} {161}
  (\bibinfo {year} {1976})}\BibitemShut {NoStop}%
\bibitem [{\citenamefont {Gray}\ and\ \citenamefont
  {Hancock}(1955)}]{Gray1955}%
  \BibitemOpen
  \bibfield  {author} {\bibinfo {author} {\bibfnamefont {J.}~\bibnamefont
  {Gray}}\ and\ \bibinfo {author} {\bibfnamefont {G.}~\bibnamefont {Hancock}},\
  }\href@noop {} {\bibfield  {journal} {\bibinfo  {journal} {Journal of
  Experimental Biology}\ }\textbf {\bibinfo {volume} {32}},\ \bibinfo {pages}
  {802} (\bibinfo {year} {1955})}\BibitemShut {NoStop}%
\bibitem [{\citenamefont {Pham}\ \emph {et~al.}(2013)\citenamefont {Pham},
  \citenamefont {Lawrence}, \citenamefont {Lee}, \citenamefont {Grason},
  \citenamefont {Emrick},\ and\ \citenamefont {Crosby}}]{Pham2013}%
  \BibitemOpen
  \bibfield  {author} {\bibinfo {author} {\bibfnamefont {J.~T.}\ \bibnamefont
  {Pham}}, \bibinfo {author} {\bibfnamefont {J.}~\bibnamefont {Lawrence}},
  \bibinfo {author} {\bibfnamefont {D.~Y.}\ \bibnamefont {Lee}}, \bibinfo
  {author} {\bibfnamefont {G.~M.}\ \bibnamefont {Grason}}, \bibinfo {author}
  {\bibfnamefont {T.}~\bibnamefont {Emrick}}, \ and\ \bibinfo {author}
  {\bibfnamefont {A.~J.}\ \bibnamefont {Crosby}},\ }\href {\doibase
  10.1002/adma.201302817} {\bibfield  {journal} {\bibinfo  {journal} {Advanced
  Materials}\ }\textbf {\bibinfo {volume} {25}},\ \bibinfo {pages} {6703}
  (\bibinfo {year} {2013})}\BibitemShut {NoStop}%
\bibitem [{\citenamefont {Lee}\ \emph {et~al.}(2013)\citenamefont {Lee},
  \citenamefont {Pham}, \citenamefont {Lawrence}, \citenamefont {Lee},
  \citenamefont {Parkos}, \citenamefont {Emrick},\ and\ \citenamefont
  {Crosby}}]{Lee2013}%
  \BibitemOpen
  \bibfield  {author} {\bibinfo {author} {\bibfnamefont {D.~Y.}\ \bibnamefont
  {Lee}}, \bibinfo {author} {\bibfnamefont {J.~T.}\ \bibnamefont {Pham}},
  \bibinfo {author} {\bibfnamefont {J.}~\bibnamefont {Lawrence}}, \bibinfo
  {author} {\bibfnamefont {C.~H.}\ \bibnamefont {Lee}}, \bibinfo {author}
  {\bibfnamefont {C.}~\bibnamefont {Parkos}}, \bibinfo {author} {\bibfnamefont
  {T.}~\bibnamefont {Emrick}}, \ and\ \bibinfo {author} {\bibfnamefont {A.~J.}\
  \bibnamefont {Crosby}},\ }\href@noop {} {\bibfield  {journal} {\bibinfo
  {journal} {Advanced Materials}\ }\textbf {\bibinfo {volume} {25}},\ \bibinfo
  {pages} {1248} (\bibinfo {year} {2013})}\BibitemShut {NoStop}%
\bibitem [{\citenamefont {Kim}\ \emph {et~al.}(2010)\citenamefont {Kim},
  \citenamefont {Lee}, \citenamefont {Sudeep}, \citenamefont {Emrick},\ and\
  \citenamefont {Crosby}}]{Kim2010a}%
  \BibitemOpen
  \bibfield  {author} {\bibinfo {author} {\bibfnamefont {H.~S.}\ \bibnamefont
  {Kim}}, \bibinfo {author} {\bibfnamefont {C.~H.}\ \bibnamefont {Lee}},
  \bibinfo {author} {\bibfnamefont {P.~K.}\ \bibnamefont {Sudeep}}, \bibinfo
  {author} {\bibfnamefont {T.}~\bibnamefont {Emrick}}, \ and\ \bibinfo {author}
  {\bibfnamefont {A.~J.}\ \bibnamefont {Crosby}},\ }\href {\doibase
  10.1002/adma.201001892} {\bibfield  {journal} {\bibinfo  {journal} {Advanced
  Materials}\ }\textbf {\bibinfo {volume} {22}},\ \bibinfo {pages} {4600}
  (\bibinfo {year} {2010})}\BibitemShut {NoStop}%
\bibitem [{\citenamefont {{Warner Jr.}}(1972)}]{WarnerJr.1972}%
  \BibitemOpen
  \bibfield  {author} {\bibinfo {author} {\bibfnamefont {H.~R.}\ \bibnamefont
  {{Warner Jr.}}},\ }\href {http://pubs.acs.org/doi/abs/10.1021/i160043a017}
  {\bibfield  {journal} {\bibinfo  {journal} {Industrial \& Engineering
  Chemistry Fundamentals}\ }\textbf {\bibinfo {volume} {1}},\ \bibinfo {pages}
  {379} (\bibinfo {year} {1972})}\BibitemShut {NoStop}%
\bibitem [{\citenamefont {Kroger}(2004)}]{Kroger2004}%
  \BibitemOpen
  \bibfield  {author} {\bibinfo {author} {\bibfnamefont {M.}~\bibnamefont
  {Kroger}},\ }\href {\doibase 10.1016/j.physrep.2003.10.014} {\bibfield
  {journal} {\bibinfo  {journal} {Physics Reports}\ }\textbf {\bibinfo {volume}
  {390}},\ \bibinfo {pages} {453} (\bibinfo {year} {2004})}\BibitemShut
  {NoStop}%
\bibitem [{\citenamefont {Zeng}\ \emph {et~al.}(2004)\citenamefont {Zeng},
  \citenamefont {Saltysiak}, \citenamefont {Johnson}, \citenamefont
  {Schiraldi},\ and\ \citenamefont {Kumar}}]{Zeng2004}%
  \BibitemOpen
  \bibfield  {author} {\bibinfo {author} {\bibfnamefont {J.}~\bibnamefont
  {Zeng}}, \bibinfo {author} {\bibfnamefont {B.}~\bibnamefont {Saltysiak}},
  \bibinfo {author} {\bibfnamefont {W.~S.}\ \bibnamefont {Johnson}}, \bibinfo
  {author} {\bibfnamefont {D.~a.}\ \bibnamefont {Schiraldi}}, \ and\ \bibinfo
  {author} {\bibfnamefont {S.}~\bibnamefont {Kumar}},\ }\href {\doibase
  10.1016/S1359-8368(03)00051-9} {\bibfield  {journal} {\bibinfo  {journal}
  {Composites Part B: Engineering}\ }\textbf {\bibinfo {volume} {35}},\
  \bibinfo {pages} {173} (\bibinfo {year} {2004})}\BibitemShut {NoStop}%
\bibitem [{\citenamefont {Pham}\ \emph {et~al.}(2014)\citenamefont {Pham},
  \citenamefont {Lawrence}, \citenamefont {Grason}, \citenamefont {Emrick},\
  and\ \citenamefont {Crosby}}]{Pham2014}%
  \BibitemOpen
  \bibfield  {author} {\bibinfo {author} {\bibfnamefont {J.~T.}\ \bibnamefont
  {Pham}}, \bibinfo {author} {\bibfnamefont {J.}~\bibnamefont {Lawrence}},
  \bibinfo {author} {\bibfnamefont {G.~M.}\ \bibnamefont {Grason}}, \bibinfo
  {author} {\bibfnamefont {T.}~\bibnamefont {Emrick}}, \ and\ \bibinfo {author}
  {\bibfnamefont {A.~J.}\ \bibnamefont {Crosby}},\ }\href {\doibase
  10.1039/c3cp55502j} {\bibfield  {journal} {\bibinfo  {journal} {Physical
  Chemistry Chemical Physics}\ }\textbf {\bibinfo {volume} {16}},\ \bibinfo
  {pages} {10261} (\bibinfo {year} {2014})}\BibitemShut {NoStop}%
\bibitem [{Note1()}]{Note1}%
  \BibitemOpen
  \bibinfo {note} {The expression of non-linearity used for this end-loading
  experiment is slightly different from the expression used for the flow
  experiments as the boundary conditions are different. $F= \protect \frac
  {4\pi ^{2}N^{2}BH}{L^{3}} \left [ \protect \frac {\protect \sqrt {1-\left (
  H_{0}/L \right )^{2}}}{\protect \sqrt {1-\left ( H/L \right )^{2}}} +M\right
  ]$ where $F$ is the force and the constant $M=2/(1+\nu )-1$ (where $\nu
  \approx 0.3$ is the Poisson's ratio). \cite {Pham2014} In the small strain
  limit where $H \ll L$, the force scales as $F\sim N^{2}BH/L^{3}$. A geometric
  relationship for a helical structure holds that $R\sim L/N$, leading to $F
  \sim BH/R^{2}L$ used to determine $B$, identical to Eq. \ref
  {endloading}.}\BibitemShut {Stop}%
\bibitem [{\citenamefont {Cross}\ and\ \citenamefont
  {Wheatland}(2012)}]{Cross2012}%
  \BibitemOpen
  \bibfield  {author} {\bibinfo {author} {\bibfnamefont {R.~C.}\ \bibnamefont
  {Cross}}\ and\ \bibinfo {author} {\bibfnamefont {M.~S.}\ \bibnamefont
  {Wheatland}},\ }\href {\doibase 10.1119/1.4750489} {\bibfield  {journal}
  {\bibinfo  {journal} {American Journal of Physics}\ }\textbf {\bibinfo
  {volume} {80}},\ \bibinfo {pages} {1051} (\bibinfo {year}
  {2012})}\BibitemShut {NoStop}%
\bibitem [{\citenamefont {Miller}\ \emph {et~al.}(2014)\citenamefont {Miller},
  \citenamefont {Lazarus}, \citenamefont {Audoly},\ and\ \citenamefont
  {Reis}}]{Miller2014}%
  \BibitemOpen
  \bibfield  {author} {\bibinfo {author} {\bibfnamefont {J.~T.}\ \bibnamefont
  {Miller}}, \bibinfo {author} {\bibfnamefont {A.}~\bibnamefont {Lazarus}},
  \bibinfo {author} {\bibfnamefont {B.}~\bibnamefont {Audoly}}, \ and\ \bibinfo
  {author} {\bibfnamefont {P.~M.}\ \bibnamefont {Reis}},\ }\href {\doibase
  10.1103/PhysRevLett.112.068103} {\bibfield  {journal} {\bibinfo  {journal}
  {Physical Review Letters}\ }\textbf {\bibinfo {volume} {112}},\ \bibinfo
  {pages} {068103} (\bibinfo {year} {2014})}\BibitemShut {NoStop}%
\bibitem [{\citenamefont {Smith}\ \emph {et~al.}(2001)\citenamefont {Smith},
  \citenamefont {Zastavker},\ and\ \citenamefont {Benedek}}]{Smith2001a}%
  \BibitemOpen
  \bibfield  {author} {\bibinfo {author} {\bibfnamefont {B.}~\bibnamefont
  {Smith}}, \bibinfo {author} {\bibfnamefont {Y.}~\bibnamefont {Zastavker}}, \
  and\ \bibinfo {author} {\bibfnamefont {G.}~\bibnamefont {Benedek}},\ }\href
  {\doibase 10.1103/PhysRevLett.87.278101} {\bibfield  {journal} {\bibinfo
  {journal} {Physical Review Letters}\ }\textbf {\bibinfo {volume} {87}},\
  \bibinfo {pages} {278101} (\bibinfo {year} {2001})}\BibitemShut {NoStop}%
\bibitem [{\citenamefont {Brochard-Wyart}(1995)}]{Brochard-Wyart1996}%
  \BibitemOpen
  \bibfield  {author} {\bibinfo {author} {\bibfnamefont {F.}~\bibnamefont
  {Brochard-Wyart}},\ }\href {\doibase 10.1209/0295-5075/30/7/002} {\bibfield
  {journal} {\bibinfo  {journal} {Europhysics Letters (EPL)}\ }\textbf
  {\bibinfo {volume} {30}},\ \bibinfo {pages} {387} (\bibinfo {year}
  {1995})}\BibitemShut {NoStop}%
\bibitem [{\citenamefont {Liu}\ and\ \citenamefont
  {Steinberg}(2010)}]{Liu2010b}%
  \BibitemOpen
  \bibfield  {author} {\bibinfo {author} {\bibfnamefont {Y.}~\bibnamefont
  {Liu}}\ and\ \bibinfo {author} {\bibfnamefont {V.}~\bibnamefont
  {Steinberg}},\ }\href@noop {} {\bibfield  {journal} {\bibinfo  {journal}
  {Europhysics Letters (EPL)}\ }\textbf {\bibinfo {volume} {90}},\ \bibinfo
  {pages} {44002} (\bibinfo {year} {2010})}\BibitemShut {NoStop}%
\end{thebibliography}%

\end{document}